\documentstyle[prd,aps,graphics]{revtex}
\begin{document}
\def\Universita{Universit\'a} 
\def\Paris{Par\'{\i}s} 
\def\Perez{P\'erez}
\def\Gunther{G\"unther} 
\def\Schutzhold{Sch\"utzhold}
\def\Lofstedt{L\"{o}fstedt} 
\def\Garcia{Garc\'\i{}a}
\def\Ruggeberg{R\"uggeberg} 
\def\SIZE{0.33}
\title{Sonoluminescence as a QED vacuum effect:
Probing Schwinger's proposal} 
\author{S. Liberati${}^{\dagger}$}
\address{International School for Advanced Studies, Via Beirut 2-4, 
34014 Trieste, Italy}
\author{Matt Visser${}^{\P}$}
\address{Physics Department, Washington University, 
Saint Louis MO 63130-4899, USA}
\author{F. Belgiorno${}^{\ddagger}$}
\address{\Universita\ degli Studi di Milano, Dipartimento di Fisica, 
Via Celoria 16,  20133 Milano, Italy}
\author{D.W. Sciama${}^{\S}$}
\address{International School for Advanced Studies, Via Beirut 2-4, 
34014 Trieste, Italy\\
International Center for Theoretical Physics,  Strada Costiera 11, 
34014 Trieste, Italy\\
Physics Department, Oxford University, Oxford, England}

\date{11 May 1998; Revised 17 June 1999; \LaTeX-ed \today}

\maketitle{}

\begin{abstract}
 
Several years ago Schwinger proposed a physical mechanism for
sonoluminescence in terms of photon production due to changes in the
properties of the quantum-electrodynamic (QED) vacuum arising from a
collapsing dielectric bubble.  This mechanism can be re-phrased in
terms of the Casimir effect and has recently been the subject of
considerable controversy. The present paper probes Schwinger's
suggestion in detail: Using the sudden approximation we calculate
Bogolubov coefficients relating the QED vacuum in the presence of the
expanded bubble to that in the presence of the collapsed bubble.  In
this way we derive an estimate for the spectrum and total energy
emitted. We verify that in the sudden approximation there is an
efficient production of photons, and further that the main
contribution to this dynamic Casimir effect comes from a volume term,
as per Schwinger's original calculation.  However, we also demonstrate
that the timescales required to implement Schwinger's original
suggestion are not physically relevant to sonoluminescence. Although
Schwinger was correct in his assertion that changes in the zero-point
energy lead to photon production, nevertheless his original model is
not appropriate for sonoluminescence. In other work we have developed
a variant of Schwinger's model that is compatible with the physically
required timescales.

\end{abstract}

\pacs{PACS:12.20.Ds; 77.22.Ch; 78.60.Mq}

PACS:12.20.Ds; 77.22.Ch; 78.60.Mq

\section{Introduction}
\def\ng{n_{gas}}
\def\nl{n_{liquid}}
\def\ni{n_{inside}}
\def\n0{n_{outside}}
\def\max{\hbox{max}}
\def\min{\hbox{min}}

In this paper we shall concentrate on Schwinger's original proposal
regarding sonoluminescence~\cite{Sc1,Sc2,Sc3,Sc4,Sc5,Sc6,Sc7}, that of
photon production associated with changes in the QED vacuum state. 
His idea was to explain the sonoluminescence phenomenon, which
consists  
in light emission by a
sound-driven gas bubble in fluid \cite{Physics-Reports}, in the   
framework of the so-called dynamical Casimir effect.
Our first aim is to verify, in a dynamic framework, that a sudden change in
bubble size will cause efficient photon production, thereby indicating
the possibility of an a priori interesting role for the dynamic
Casimir effect in this condensed matter context. While we demonstrate
that the key features of Schwinger's calculations are correct, this
study also demonstrates that for other reasons (to do with the
observed timescale of the phenomenon) the original approach of
Schwinger is not physically relevant to sonoluminescence. In related
work~\cite{letter,qed1,qed2,2gamma} we have developed a different
implementation of Schwinger's ideas regarding sonoluminescence that is
compatible with the physically observed timescales.

The idea of a ``Casimir route'' to sonoluminescence 
was developed by Schwinger in a 
series of papers~\cite{Sc1,Sc2,Sc3,Sc4,Sc5,Sc6,Sc7}. 
One key issue in Schwinger's model is
simply that of calculating static Casimir energies for dielectric
spheres---and there is already considerable disagreement on this
issue. A second and in some ways more critical question is the extent
to which a change in static Casimir energies might be converted to
real photons during the collapse of the bubble---it is this issue that
we shall address in this paper. We estimate the spectrum of the
emitted photons by calculating an appropriate Bogolubov coefficient
relating the two states of the QED vacuum.  

Another model associating sonoluminescence with QED vacuum changes is
the variant of Schwinger's proposal due to
Eberlein~\cite{Eberlein1,Eberlein2,Eberlein3}.  In contrast to
Schwinger's quasi-static approach, Eberlein's model is truly dynamical
but uses a radically different physical approximation---the adiabatic
approximation.  The two models should not be confused. See~\cite{qed1}
for a deeper discussion of Eberlein's approach to sonoluminescence.

Considerable confusion has been caused by Schwinger's choice of the
phrase ``dynamical Casimir effect'' to describe his model.  In fact,
the original model is not dynamical and is better described as
quasi-static as the heart of the model lies in comparing two static
Casimir energy calculations: that for an expanded bubble with that for
a collapsed bubble.  In a series of
papers~\cite{Sc1,Sc2,Sc3,Sc4,Sc5,Sc6,Sc7} Schwinger showed that the
dominant bulk contribution to the Casimir energy of a bubble (of
dielectric constant $\epsilon_{inside}$) in a dielectric background
(of dielectric constant $\epsilon_{outside}$) is
\begin{eqnarray}
E_{cavity}
&=&
+2\frac{4\pi}{3}R^3 \; \int_0^K
{4 \pi k^2 dk\over(2\pi)^3}    \; \frac{1}{2}  \hbar c k
\left(
{1\over\sqrt{\epsilon_{inside}}}- {1\over\sqrt{\epsilon_{outside}}}
\right) +\cdots
\nonumber\\
&=&+\frac{1}{6 \pi} \hbar c R^3 K^4
\left(
{1\over\sqrt{\epsilon_{inside}}}- {1\over\sqrt{\epsilon_{outside}}}
\right) +\cdots.
\label{Esch}
\end{eqnarray}
There are additional sub-dominant finite volume
effects~\cite{CMMV1,CMMV2,MV}. The quantity $K$ is a high-wavenumber
cutoff that characterizes the wavenumber at which the dielectric
constants drop to their vacuum values. This result can also be
rephrased in the clearer and more general form as~\cite{CMMV1,CMMV2,MV}:
\begin{equation}
E_{ cavity} = + 2 V \int \frac{d^3\vec{k}}{(2 \pi)^3} \; \frac{1}{2}
\hbar  \left[ \omega_{inside}(k) - \omega_{outside}(k)  \right] + \cdots
\end{equation}
where it is evident that the Casimir energy can be interpreted as
a difference in zero point energies due to the different dispersion
relations inside and outside the bubble.

In contrast, Milton~\cite{M95}, and Milton and Ng~\cite{M96,M97}
strongly criticize Schwinger's result.  Using what is to our minds a
physically dubious renormalization argument leads them~\cite{M97} to
discard both the volume and even the surface term and to claim that
the Casimir energy for any dielectric bubble is of order $E\approx
\hbar c/R$.

In~\cite{CMMV1,CMMV2,MV} an extensive discussion on these topics is
found.  Therein it is emphasized that one has to compare two different
geometrical configurations, and different quantum states, of the same
spacetime regions. In a situation like that of Schwinger's model one
has to subtract from the zero point energy (ZPE) for a vacuum bubble
in water the ZPE for water filling all space. It is clear that in this
case the bulk term is physical and {\em must} be taken into
account. In the situation pertinent to sonoluminescence, the total
volume occupied by the gas is not constant (the gas is truly
compressed), and it is far too naive to simply view the ingoing water
as flowing coherently from infinity (leaving voids filled with air or
vacuum somewhere in the apparatus). Since the density of water is
approximately but not exactly constant, the influx of water will
instead generate an outgoing density wave which will be rapidly damped
by the viscosity of the fluid. The few phonons generated in this way
are surely negligible. Surface terms are also present,
and eventually other higher order correction terms, but they prove
to not be dominant for sufficiently large cavities~\cite{MV}.

\section{Bogolubov coefficients}

As a first approach to the problem we study in detail the basic
mechanism of particle creation, and test the consistency of the
Casimir energy proposals previously described.  With this aim in mind
we consider the change in the QED vacuum associated with the
collapse of the bubble, by keeping fixed the refractive index both of
the gas and of the water.  For the sake of simplicity we take, as
Schwinger did, only the electric part of QED, reducing the problem to
a scalar electrodynamics.  Moreover, at this stage of development, we
are not concerned with the dynamics of the bubble surface.  In analogy
with the subtraction procedure of the quasi-static calculations of
Schwinger~\cite{Sc1,Sc2,Sc3,Sc4,Sc5,Sc6,Sc7}, and of Carlson et
al.~\cite{CMMV1,CMMV2,MV}, we shall consider two different
configurations of space.  An ``in'' configuration with a bubble of
dielectric constant $\epsilon_{inside}$ (typically vacuum) in a medium
of dielectric constant $\epsilon_{outside}$, and an ``out'' one in
which one has just the latter medium (dielectric constant
$\epsilon_{outside}$) filling all space.  Strictly speaking we should
compare a large bubble having radius $R_{\mathrm max}$ with a small
bubble of radius $R_{\mathrm min}$. We are approximating the small
bubble by zero volume on the grounds that the small bubble that is
relevant to sonoluminescence is at least a million times smaller than
the large bubble at the expansion maximum. Keeping $R_{\mathrm min}$
finite significantly complicates the calculation but does not give
much more physical information. The above ``in'' and ``out''
configurations will correspond to two different bases for the
quantization of the field.  The two bases will be related by Bogolubov
coefficients in the usual way. Once we determine these coefficients we
easily get the number of created particles per mode and from this the
spectrum. This tacitly makes the ``sudden approximation'': Changes in
the refractive index are assumed to be non-adiabatic,
see~\cite{letter,qed1,qed2} for more discussion. We shall also make a
consistency check by a direct confrontation between the change in
Casimir energy and the direct sum, $E=\sum_{k} \omega_{k} n_{k}$ of
the energies of the created photons.  The former energy (the total
energy of the particles that can be produced by the collapse) must
necessarily equal the Casimir energy of the bubble in the ``in'' state
since in the current simplified model there is no external source of
energy (like the driving sound in the true dynamical effect). For this
reason we expect to be able to give a definitive answer on the nature
(dependence on the bubble radius and on the cut-off) of the static
Casimir energy.  Of course it is evident that such a model cannot be
considered a fully satisfactory model for sonoluminescence. In fact
this model completely ignores the details of the dynamics and
moreover, by considering just one cycle, implies impossibility of
testing for the possible presence of any parametric resonances.  We
thus consider the present calculation as a toy model in which some
basic features of the Casimir approach to sonoluminescence are
investigated: it provides a test of the nature and quantity of the
particles produced by a collapsing dielectric bubble in the sudden
approximation.

\subsection{Formal calculation}

Let us consider the equations of the electric fields (Schwinger
framework) in spherical coordinates and with a time independent
dielectric constant (we temporarily set $c=1$ for ease of notation,
and shall reintroduce appropriate factors of the speed of light
when needed for clarity)
\begin{equation}
\epsilon \partial_{0}(\partial_{0} E)-\nabla^{2} E=0.
\end{equation}
We look for solutions of the form
\begin{equation}
E=\Phi(r,t)Y_{lm}(\Omega) {1\over r}.
\end{equation}
Then one finds
\begin{equation}
\epsilon(\partial^{2}_{0}\Phi)-(\partial_{r}^{2} \Phi)+{1\over r^{2}}
l(l+1) \Phi =0.
\label{E:eqm2}
\end{equation}
For both the ``in'' and ``out'' solution the field equation in $r$
is given by:
\begin{equation}
\epsilon\partial_{0}^{2} \Phi-\partial_{r}^{2}\Phi +{1\over r^{2}} l(l+1)
\Phi=0.
\end{equation}
In both asymptotic regimes (past and future) one has a static
situation (either a bubble in the dielectric, or just the dielectric) so
one can in this limit factorize the time and radius dependence of
the modes:  $\Phi(r,t)=e^{i\omega t} f(r)$.  One gets
\begin{equation}
f^{''}+\left (\epsilon \omega^{2} -{1\over r^{2}} l(l+1)\right) f=0.
\end{equation}
This is a well known differential equation. To handle it more easily
in a standard way we can cast it as an eigenvalues problem
\begin{equation}
f^{''}-\left( {1\over r^{2}} l(l+1) \right)f=-\lambda^{2}f,
\end{equation}
where $\lambda^2=\epsilon \omega^{2}$.
With the change of variables $f=r^{1/2} G$ we get
\begin{equation}
G^{''}+{1\over r}G^{'}+\left(\lambda^{2}-{\nu^2 \over r^{2}} \right)G=0.
\end{equation}
This is the standard Bessel equation. It admits as solutions the
first type Bessel and Neumann functions, $J_{\nu}(\lambda r)$ and
$N_{\nu}(\lambda r)$, with $\nu=l+1/2$.  Remember that for those
solutions which have to be well-defined at the origin, $r=0$,
regularity implies the absence of the Neumann functions.  For the
``in'' basis we have to take into account that the dielectric
constant changes at the bubble radius ($R$). In fact we have
\begin{equation}
\epsilon=\left \{ 
\begin{array}{llllll}
\epsilon_{inside} & = & n^2_{gas} & = & 
\mbox{dielectric constant of air-gas mixture} & 
\mbox{if $r\leq R$},\\
\epsilon_{outside} & = & n^2_{liquid} & = & 
\mbox{dielectric constant of ambient liquid (typically water)} & 
\mbox{if $r > R$}.
\end{array}
\right.
\end{equation}
Typically one simplifies calculations by using the fact that the
dielectric constant of air is approximately equal 1 at standard
temperature and pressure (STP),  and then dealing only with the
dielectric constant of water ($n_{liquid} = \sqrt{\epsilon_{outside}}
\approx 1.3$). We find it convenient to explicitly keep track of 
$n_{gas}$ and $n_{liquid}$ in the formalism we develop.  Defining the
in and out frequencies, $\omega_{in}$ and $\omega_{out}$
respectively, one has
\begin{equation}
G^{in}_{\nu}(\ng,\nl,\omega_{in},r)=\left \{
\begin{array}{ll}
A_{\nu} J_{\nu}(\ng\, \omega_{in} r) & \mbox{if $r\leq R$},\\ 
B_{\nu} J_{\nu}(\nl\, \omega_{in} r)+
C_{\nu} N_{\nu}(\nl\, \omega_{in} r)&
\mbox{if $r > R$.}
\end{array}
\right.
\end{equation}
The $A_{\nu}$, $B_{\nu}$, and $C_{\nu}$ coefficients are determined
by matching conditions in $R$
\begin{equation}
\begin{array}{lll}
A_{\nu} J_{\nu}(\ng\, \omega_{in}  R)&=&
B_{\nu} J_{\nu}(\nl\, \omega_{in} R)+C_{\nu}N_{\nu}(\nl\, \omega_{in} R),\\
A_{\nu} J_{\nu}{'}(\ng\, \omega_{in} R)&=&
B_{\nu} J_{\nu}{'}(\nl\, \omega_{in} R)+C_{\nu}
N_{\nu}{'}(\nl\, \omega_{in} R).
\end{array}
\label{E:coef}
\end{equation}
The ``out'' basis is easily obtained solving the same equation but for a
space filled with a homogeneous dielectric,
\begin{equation}
G^{out}_{\nu}(\nl,\omega_{out},r) = J_{\nu}(\nl\, \omega_{out} r).
\end{equation}
To check that the ``out'' basis is properly normalized we use the
scalar product, defined as usual by
\begin{equation} 
(\phi_{1},\phi_{2})=-i \int_{\Sigma}\phi_{1}
\stackrel{\leftrightarrow}{\partial}_{0}\phi_{2}^{*} \: d^{3}x.
\end{equation}
There are subtleties in the definition of scalar product which are
dealt with more fully in~\cite{letter,qed1,qed2}.  The naive scalar
product adopted here is missing a dependence on the refractive
indices of the gas and the surrounding water.  Given the fact that in
the present framework both of these are approximately equal to one,
the product adopted here is good enough for a qualitative discussion.
Consider now the scalar product of a eigenfunction with itself, one
expects to obtain a normalization condition which can be written as
\begin{equation}
((\Phi^{i}_{out})^{*},(\Phi^{j}_{out})^{*})=\delta^{ij}.
\end{equation}
Inserting the explicit form of the $\Phi$ functions 
we get 
\begin{eqnarray}
((\Phi^{i}_{out})^{*},(\Phi^{j}_{out})^{*})&=&
(\lambda+\lambda^{\prime}) \int_{0}^{\infty} r dr 
J_{\nu}(\lambda r)J_{\nu}(\lambda^{\prime} r) 
e^{i(\lambda-\lambda^{\prime})}\\
&=& (\lambda+\lambda^{\prime}) 
\frac{\delta(\lambda-\lambda^{\prime})}{\lambda} 
e^{i(\lambda-\lambda^{\prime})},  
\end{eqnarray} 
where we have used the Hankel Integral Formula \cite{Bateman}
\begin{equation}
\int_{0}^{\infty}   r dr J_{\nu}(\lambda r)J_{\nu}(\lambda^{\prime} 
r)=\delta(\lambda-\lambda^{\prime})/\lambda. 
\end{equation}
The Bogolubov coefficients are {\em defined} as
\begin{eqnarray}
\alpha_{ij}&=&-({E_{i}^{out}}^{*},{E_{j}^{in}}^{*}),\\
\beta_{ij} &=&+(E_{i}^{out}, {E_{j}^{in}}^{*}).
\end{eqnarray}
We are mainly interested in the coefficient $\beta$, since
$|\beta|^{2}$ is linked to the total number of particles created.
By a direct substitution it is easy to find the expression:
\begin{eqnarray}
\beta &=&-i \int_{0}^{\infty} 
\left( \Phi_{out}(r,t) \; Y_{lm}(\Omega)  \; {1\over r} \right) 
\stackrel{\leftrightarrow}{\partial}_{0}
\left( \Phi_{in}(r,t) \; Y_{l^{\prime}m^{\prime}}(\Omega)  \;{1\over r}
\right)^{*}\: r^2 dr d\Omega,\\
&=& (\omega_{in}-\omega_{out}) \; e^{i(\omega_{out}+\omega_{in})t} 
\delta_{l l^{\prime}}\; \delta_{m m^{\prime}} \; 
\int_{0}^{\infty}G^{out}_{l}(\nl,\omega_{out},r) \; 
G^{in}_{l^{\prime}}(\ng,\nl,\omega_{in},r) \: r dr.
\label{E:beta1}
\end{eqnarray}
To compute the integral one needs some ingenuity. Let us write the
equations of motion for two different values of the eigenvalues,
$\lambda$ and $\mu$:
\begin{eqnarray}
G_{\lambda}^{''}+{1\over r}G_{\lambda}^{'}+\left (\lambda^{2}-
{1\over r^{2}} (l+{1\over 2})^2  \right)G_{\lambda}&=&0,\\
G_{\mu}^{''}+{1\over r}G_{\mu}^{'}+\left (\mu^{2}-
{1\over r^{2}} (l+{1 \over 2})^2 \right)G_{\mu}&=&0.
\end{eqnarray}
If we multiply the first by $G_{\mu}$ and
the second by $G_{\lambda}$ we get
\begin{eqnarray}
G_{\lambda}^{''}G_{\mu}+{1\over r}G_{\lambda}^{'}G_{\mu}+ \left
(\lambda^{2}-
{1\over r^{2}} (l+{1\over 2})^2\right)G_{\lambda}G_{\mu}&=&0,\\
G_{\mu}^{''}G_{\lambda}+{1\over r}G_{\mu}^{'}G_{\lambda}+ \left (\mu^{2}-
{1\over r^{2}} (l+{1\over 2})^2 \right)G_{\mu}G_{\lambda}&=&0.
\end{eqnarray}
Subtracting the second from the first we then obtain
\begin{equation}
\left( G_{\lambda}^{''}G_{\mu}-G_{\mu}^{''}G_{\lambda}\right)+
{1\over r}\left(G_{\lambda}^{'}G_{\mu}-G_{\mu}^{'}G_{\lambda}\right)+
(\lambda^{2}-\mu^{2}) G_{\lambda}G_{\mu}=0.
\end{equation}
The second term on the left hand side is a pseudo-Wronskian
determinant
\begin{equation}
W_{\lambda\mu}(r)
=
G_{\lambda}^{'}(r) G_{\mu}(r) -G_{\mu}^{'}(r) G_{\lambda}(r),
\end{equation}
and the first term is its total derivative $dW_{\lambda\mu}/dr$.
It's a pseudo-Wronskian, not a true Wronskian, since the two
functions $G_{\lambda}$ and $G_{\mu}$ correspond to different
eigenvalues and so solve different differential equations. The
derivatives are all with respect to the variable $r$.  Using this
definition we can cast the integral over $r$ of the product of two
given solutions into a simple form. Generically:
\begin{equation}
(\mu^{2}-\lambda^{2}) \int_{a}^{b} r dr\; G_{\lambda}G_{\mu}
=
\int_{a}^{b} rdr \; dW_{\lambda \mu}+ \int_{a}^{b} 
dr\; W_{\lambda \mu}.
\end{equation}
That is
\begin{equation}
\int_{a}^{b} r dr\; G_{\lambda}G_{\mu}
=
\left. {1\over
(\mu^{2}-\lambda^{2}) } W_{\lambda \mu}\:r\right|_{a}^{b} 
-\int_{a}^{b} dr \;
W_{\lambda \mu}+ \int_{a}^{b} dr \; W_{\lambda \mu}.
\end{equation}
So the final result is
\begin{equation}
\int_{a}^{b} r \; dr \; G_{\lambda}G_{\mu}= 
\left.
{1\over (\mu^{2}-\lambda^{2}) } \; \left( W_{\lambda \mu} \; r\right) 
\right|^{b}_{a}.
\end{equation}

This expression can be applied in our specific case equation
(\ref{E:beta1}), we obtain:
\begin{eqnarray}
\int^{\infty}_{0} &r& \; dr \: 
G^{out}_{\nu}(\nl,\omega,r) \; G^{in}_{\nu}(\ng,\nl,\omega,r)
\nonumber\\
&=&\int^{R}_{0} r\; dr\: 
G^{out}_{\nu}(\nl\,\omega_{out}r)G^{in}_{\nu}(\ng\,\omega_{in}r)
+\int_{R}^{\infty} r \; dr \: G^{out}_{\nu}(\nl\,\omega_{out}r) 
G^{in}_{\nu}(\nl\,\omega_{in}r)\\
&=& \frac{
\left\{ r 
W[G^{out}_{\nu}(\nl\,\omega_{out}r),G^{in}_{\nu}(\ng\,\omega_{in}r)]
\right\}^{R}_{0}
}
{(\nl\,\omega_{out})^2-(\ng\,\omega_{in})^2} +
\frac{
\left\{r 
W[G^{out}_{\nu}(\nl\,\omega_{out}r),G^{in}_{\nu}(\nl\,\omega_{in}r)]
\right\}^{\infty}_{0}
}
{(\nl\,\omega_{out})^2-(\nl\,\omega_{in})^2}\\
&=& R\,
\left[
\frac{
W[G^{out}_{\nu}(\nl\,\omega_{out}r),G^{in}_{\nu}(\ng\,\omega_{in}r)]_{R_{-}}
}
{(\nl\,\omega_{out})^2-(\ng\,\omega_{in})^2}-
\frac{
W[G^{out}_{\nu}(\nl\,\omega_{out}r),G^{in}_{\nu}(\nl\,\omega_{in}r)]_{R_{+}}
}
{(\nl\,\omega_{out})^2-(\nl\,\omega_{in})^2}
\right],
\end{eqnarray}
where we have used the fact that the above forms are well behaved
(and equal to $0$) for $r=0$, and $r=\infty$ by construction. (Here
and henceforth we shall automatically give the same $l$ value to
the ``in'' and ``out'' solutions by using the fact that  equation
(\ref{E:beta1}) contains a Kronecker delta in $l$ and $l^{\prime}$.)

Finally the two pseudo-Wronskians so found can be shown to be equal
(by the junction condition (\ref{E:coef})). In fact one can easily
check that:
\begin{eqnarray}
A_{\nu}W[J_{\nu}(\nl\,\omega_{out}r),J_{\nu}(\ng\,\omega_{in}r)]_{R}
&=&
B_{\nu}W[J_{\nu}(\nl\,\omega_{out}r),J_{\nu}(\ng\,\omega_{in}r)]_{R}
\nonumber\\
&+&
C_{\nu}W[J_{\nu}(\nl\,\omega_{out}r),N_{\nu}(\ng\,\omega_{in}r)]_{R}.
\end{eqnarray}
This equality allows to rewrite integral in equation (\ref{E:beta1})
in a more compact form
\begin{eqnarray}
\int^{\infty}_{0} &r& \; dr \:
G^{out}_{\nu}(\nl\,,\omega,r) \; G^{in}_{\nu}(\ng,\nl,\omega,r)
\nonumber\\
&=& A_{\nu}
\left[ 
\frac{1}{(\nl\,\omega_{out})^2-(\ng\,\omega_{in})^2}-
\frac{1}{(\nl\,\omega_{out})^2-(\nl\,\omega_{in})^2}
\right]
W[J_{\nu}(\nl\,\omega_{out}r),J_{\nu}(\ng\,\omega_{in}r)]_{R}\\
&=&-\left(\frac{\nl^2-\ng^2}{\nl^2}\right) 
{A_{\nu} R \omega_{in}^2 \over [\omega_{out}^2-\omega_{in}^2]}\:
\frac{
W[J_{\nu}(\nl\,\omega_{out}r),J_{\nu}(\ng\,\omega_{in}r)]_{R}
}
{[(\nl\,\omega_{out})^2-(\ng\,\omega_{in})^2]}.
\end{eqnarray}
Inserting this expression into equation (\ref{E:beta1}) we get
\begin{equation}
\beta=\left(\frac{\nl^2-\ng^2}{\nl^2}\right) 
\delta_{l l^{\prime}} \;
\delta_{m m^{\prime}}
\frac{(\omega_{out}-\omega_{in})}{\omega_{out}^2-\omega_{in}^2} R A_\nu \:
\frac{\omega_{in}^2
W[J_{\nu}(\nl\,\omega_{out}r),J_{\nu}(\ng\,\omega_{in}r)]_{R}} 
{[(\nl\,\omega_{out})^2-(\ng\,\omega_{in})^2]}
e^{i(\omega_{out}+\omega_{in})t}.  
\end{equation}
We are mainly interested in the square of this coefficient summed
over $l$ and $m$. It is in fact this quantity that is linked to
the spectrum of the ``out'' particles present in the ``in'' vacuum,
and it is this quantity that is related to the total energy emitted.
Including all appropriate dimensional factors ($c$, $\hbar$) we
have
\begin{equation}
{dN(\omega_{out})\over d\omega_{out}} 
=\left(\int|\beta(\omega_{in},\omega_{out})|^{2} 
d\omega_{in} \right),
\end{equation}
and
\begin{equation}
E= \hbar \int {dN(\omega_{out})\over d\omega_{out}} 
\omega_{out} d\omega_{out}.
\end{equation}
Hence we shall deal with the computation of:
\begin{eqnarray}
\left|\beta(\omega_{in},\omega_{out})\right|^{2}
&=& \sum_{lm}\sum_{l^{\prime}m^{\prime}}
\left[\beta_{lm,l^{\prime}m^{\prime}}(\omega_{in},\omega_{out})\right]^2
\\
&=& \left(\frac{\nl^2-\ng^2}{\nl^2}\right)^2
    \left(\frac{\omega_{in}^2 R}{\omega_{out}+\omega_{in}}\right)^2 
\sum_{l=1}^\infty (2l+1) 
 \left| A_{\nu} \right|^{2}
 \left[ 
        {W[J_{\nu}(\nl\,\omega_{out}r/c),J_{\nu}(\ng\,\omega_{in}r/c)]_{R}
        \over
        (\nl\,\omega_{out})^2-(\ng\,\omega_{in})^2} 
  \right]^2.
\label{E:b2}
\end{eqnarray}
This expression is too complex to allow an analytical resolution
of the problem. Nevertheless we shall show that the terms appearing
in it can be suitably approximated in such a way as to obtain a
computable form that shall give us some information about the main
predictions of this model.  We shall first look at the large volume
limit, which will allow us to compare this result to Schwinger's
calculation, and then develop some numerical approximations suitable
to estimating the predicted spectra for finite volume.

\subsection{Behaviour in the large $R$ limit}

One of the main objectives of this calculation is to shed some light
on the extent to which the change in static Casimir energy can be
transformed into photons.  In particular we expect that the total
energy of the photons calculated from this Bogolubov approach would
give approximately the same results as the static Casimir energy
calculations such those of Schwinger, and of Carlson {\em et al.}
\cite{Sc4,M95,CMMV1}, since we have excluded any external forces.

{From} equation (\ref{E:b2}) it is easy to check that the general form
of the squared Bogolubov coefficient is given by
\begin{equation}
|\beta(x,y)|^{2}=\frac{R^2}{c^2}\beta_{0}^{2}(x,y),
\end{equation}
where $\beta_{0}^{2}(x,y)$ is a dimensionless quantity and we
introduce dimensionless variables $x=\nl\, \omega_{out} R/c$ and
$y=\ng\, \omega_{in} R/c$.  (The dimensions of $\beta$ should always
be those of time.) The spectrum is then given by
\begin{equation}
{dN(\omega_{out})\over d\omega_{out}}
=\frac{R}{c\;\ng}\left(\int_{0}^{R K}|\beta_0(x,y)|^{2}dy\right), 
\label{E:N}
\end{equation}
and the energy radiated is
\begin{equation}
E=\frac{\hbar c}{\nl^{2}\,\ng\, R} 
\int_{0}^{\infty}dx\int_{0}^{R K} dy \;
x \; |\beta_{0}(x,y)|^{2}.
\label{E:E}
\end{equation}
If $R$ is very large (but finite in order to avoid infra-red
divergences) then the ``in" and the ``out" modes can both be
described by ordinary Bessel functions
\begin{eqnarray}
G^{in}(\ng,\omega,r)&=&J_{\nu}(\ng\,\omega_{in} r/c),\\
G^{out}(\nl,\omega,r)&=&J_{\nu}(\nl\,\omega_{out} r/c).
\end{eqnarray}
We can now compute the Bogolubov coefficient relating these states
\begin{eqnarray}
\beta_{ij}&=&(E^{out}_{i},{E^{in}_{j}}^{*})\\
&=& {(\omega_{in}-\omega_{out})\over c^2} 
e^{i(\omega_{out}+\omega_{in})t}
\; \delta_{ll^{\prime}} 
\; \delta_{mm^{\prime}} 
\; \int J_{\nu}(\ng\,\omega_{in} r/c)
J_{\nu}(\nl\,\omega_{out} r/c) \; r\: dr\\
&=&(\omega_{in}-\omega_{out})e^{i(\omega_{out}+\omega_{in})t} 
\; \delta_{ll^{\prime}} 
\; \delta_{mm^{\prime}}
\; \frac{\delta(\ng\,\omega_{in}-\nl\,\omega_{out})}{\ng\,\omega_{in}}\\
&=&\left( {1\over\ng} - {1\over\nl}\right) 
e^{i\omega_{in}(\ng/\nl+1)t} 
\; \delta_{ll^{\prime}}
\; \delta_{mm^{\prime}} 
\; \delta(\ng\,\omega_{in}-\nl\,\omega_{out}).
\end{eqnarray}
This result implies that
\begin{eqnarray}
|\beta(\omega_{in},\omega_{out})|^{2}&=&\sum_{lm}
\sum_{l^{\prime}m^{\prime}}|\beta_{lml^{\prime}m^{\prime}}
(\omega_{in},\omega_{out})|^{2}\\
&=&\left({\nl-\ng\over\nl\,\ng}\right)^{2}
\sum_{l}(2l+1) \; \{ \delta(\ng\,\omega_{in}-\nl\,\omega_{out})\}^2\\
&=&\left({\nl-\ng\over\nl\,\ng}\right)^{2}
\sum_{l}(2l+1)\; \delta(0)\; \delta(\ng\,\omega_{in}-\nl\,\omega_{out})\\
&=&\left({\nl-\ng\over\nl\,\ng}\right)^{2}
\sum_{l}(2l+1)\; {R\over2\pi c}\; 
\delta(\ng\,\omega_{in}-\nl\,\omega_{out}),
\end{eqnarray}
where we have invoked the standard scattering theory result
\begin{equation}
\{\delta^{3}(k)\}^{2}={V\over(2\pi)^3} \delta^{3}(k),
\end{equation}
specialized to the fact that we have a 1-dimensional delta function
(in frequency, not momentum).  The sum over angular momenta (which
is formally infinite)  can now be estimated as follows
\begin{equation}
\sum_{l=1}^{l_{max}(\omega_{out})}(2l+1)=l_{max}^{2}(\omega_{out})-1 
\approx 
l_{max}^{2}(\omega_{out}).
\end{equation}
The angular momentum cutoff is estimated by taking
\begin{equation}
l_{max}(\omega) 
\approx {\left( \nl\, \hbar \omega_{out} /c \right) \times R \over \hbar} 
= \nl\, \omega_{out} R/c = x.
\end{equation}
So in the above we are justified in approximating
\begin{equation}
\sum_{l}(2l+1) \approx x^2.
\end{equation}
By changing to the dimensionless variables $(x,y)$ this finally gives
\begin{eqnarray}
|\beta(x,y)|^{2}&=&\left({\nl-\ng\over\nl\,\ng}\right)^{2} 
\frac{R^2}{2\pi c^2} \; x^{2} \;
\delta(x-y).
\label{E:largeb}
\label{E:infinite-volume}
\end{eqnarray}
We can now compute the spectrum and the total energy of the emitted
photons
\begin{eqnarray}
{dN(\omega_{out})\over d\omega_{out}}&\approx& 
\frac{R}{2\pi c\, \ng} \left({\nl-\ng\over\nl\,\ng}\right)^{2}
\int_{0}^{R K}\; x^{2}\; \delta(x-y) \; dy \nonumber \\ 
&=& \frac{R}{2\pi c\, \ng} 
\; \left({\nl-\ng\over\nl\,\ng}\right)^{2}x^{2} 
\; \Theta(R K -x)\nonumber\\
&=& {\nl^{2}\over \ng} 
\; \left({\nl-\ng\over\nl\,\ng}\right)^{2}
\; \frac{R^{3}\omega_{out}^{2}}{2\pi c^{3}}\; 
\Theta(K-\nl\,\omega_{out}/c),
\end{eqnarray}
For future comparison purposes it is convenient to write the spectrum
in dimensionless form as
\begin{eqnarray}
\label{E:dimensionless-spectrum-1}
{dN\over dx}&\approx& 
\frac{1}{2\pi \; \nl \; \ng} 
\; \left({\nl-\ng\over\nl\,\ng}\right)^{2} x^{2} 
\; \Theta(R K -x).
\end{eqnarray}
>From this equation it is easy to get the total number of the created
photons:
\begin{eqnarray}
\label{Number}
N&\approx& 
\frac{1}{2\pi \; \nl \; \ng} 
\; \left({\nl-\ng\over\nl\,\ng}\right)^{2} 
\int_{0}^{\infty}
x^{2}  \; \Theta(R K -x)\;dx \nonumber\\
&=&  \frac{1}{6\pi \; \nl \; \ng} 
\; \left({\nl-\ng\over\nl\,\ng}\right)^{2} \; \left(R K\right)^3.
\end{eqnarray}
and the total emitted energy
\begin{eqnarray}
E&\approx&
\frac{\hbar c}{2\pi \; \nl^{2} \, \ng \, R}
\; \left({\nl-\ng\over\nl\,\ng}\right)^{2} \int_{0}^{\infty} dx 
\int_{0}^{R K} x\times x^{2} \times \delta(x-y) dy \nonumber
\\
&=&
\frac{\hbar c}{2\pi \; \nl^{2} \, \ng \, R}
\; \left({\nl-\ng\over\nl\,\ng}\right)^{2} \int_{0}^{RK} dx  x^3
\nonumber 
\\
&=&\frac{\hbar c}{2\pi \, \nl^{2} \, \ng R} 
\; \left({\nl-\ng\over\nl\,\ng}\right)^{2} 
\frac{(R K)^{4}}{4} \nonumber\\
&=& \frac{1}{8\pi\, \nl^{2}\, \ng} 
\; \left({\nl-\ng\over\nl\,\ng}\right)^{2} 
\hbar c K \; (R K)^{3}. 
\label{energy}
\end{eqnarray}
Hence, feeding our results (\ref{E:largeb}) into equations (\ref{E:N})
and (\ref{E:E}) for $dN(\omega)/d\omega$ and $E$, we find a result
which is in substantial agreement with the Schwinger (and Carlson {\em
et al.}) results. We view this is definitive proof that indeed
Schwinger was essentially correct: The main contribution to the
Casimir energy which can be extracted from the collapse of a (large)
dielectric bubble is a bulk effect. The total energy radiated in
photons balances the change in the Casimir energy up to factors of
order one which the present analysis is too crude to detect. (For
infinite volume the whole calculation can be re-phrased in terms of
plane waves to accurately fix the last few prefactors.)

In Schwinger's original model he took $\ng\approx 1$, $\nl \approx
1.3$, $R \approx R_{\mathrm max} \approx 40 \; \mu {\rm m}$ and
$K\approx 2\pi/(360\; {\rm nm})$ \cite{Sc4}.  Then $KR \approx
698$. Substitution of these numbers into equation (\ref{Esch}) leads
to an energy budget suitable for about {\em three} million emitted
photons.

By direct substitution in equation (\ref{Number}) it is easy to check
that Schwinger's results can qualitatively be recovered also in our
formalism: in our case we get about {\em 0.7 million} photons for the
same numbers of Schwinger and about {\em 1.5 million} photons using
the updated experimental figures $R_{\mathrm max} \approx 45 \;
\mu {\rm m}$ and  $K\approx 2\pi/(300\; {\rm nm})$.  

The average energy per emitted photon is approximately\footnote{%
The maximum photon energy is $\hbar \; \omega_{\mathrm max} \approx 4\;
{\rm eV}$.  }
%
\begin{equation}
\langle E \rangle =
{3\over 4} \hbar c\; K/\nl = {3\over 4} \hbar  \; \omega_{\mathrm
max}\sim 3 \; {\rm eV}.
\end{equation}
                                      
It is important to stress that equation~(\ref{Esch}) and
equation~(\ref{energy}) are not identical (even if in the large $R$
limit the leading term of Casimir energy of the ``in'' state and the
total photon energy coincide). One can easy see that the volume term
we just found [equation~(\ref{energy})] is of second order in $(n-1)$
and not of first order like equation~(\ref{Esch}).  This is ultimately
due to the fact that the interaction term responsible for converting
the initial energy in photons is a pairwise squeezing
operator~\cite{2gamma}.  Equation (\ref{energy}) demonstrates that any
argument that attempts to deny the relevance of volume terms to
sonoluminescence due to their dependence on $(n-1)$ has to be
carefully reassessed.  In fact what you measure when the refractive
index in a given volume of space changes is {\em not} directly the
static Casimir energy of the ``in'' state, but rather the fraction of
this static Casimir energy that is converted into photons. We have
just seen that once conversion efficiencies are taken into account,
the volume dependence is conserved, but not the power in the
difference of the refractive index. 

Indeed the dependence of $|\beta|^2$ on $(n-1)^2$ and the symmetry of
the former under the interchange of ``in'' and ``out'' state also
proves that it is the amount of change in the refractive index and not
its ``direction'' (from ``in'' to ``out'') that governs particle
production. This also implies that any argument using static Casimir
energy balance over a full cycle has to be used very carefully.
Actually the total change of the Casimir energy of the bubble over a
cycle would be zero (if the final refractive index of the gas is again
1).  Nevertheless in the dynamical calculation one gets photon
production in both collapse as well expansion phases. (Although some
destructive interferences between the photons produced in collapse and
in expansion are conceivable, these will not be really effective in
depleting photon production because of the substantial dynamical
difference between the two phases and because of the, easy to check,
fact that most of the photons created in the collapse will be far away
from the emission zone by the time the expansion photons would be
created.)  This apparent paradox is easily solved by taking into
account that the main source of energy is the sound field and that the
amount of this energy actually converted in photons during each cycle
is a very tiny amount of the total power.

We now turn to the study of the predictions of the model in the
case of finite radius. Unfortunately this cannot be done in an
analytic way due to the wild behaviour of the pseudo-Wronskian of
the Bessel function.  Nevertheless some ingenuity and a detailed
study of the different parts of the Bogolubov coefficient leads
to reasonable approximations that allow a clear description of the
spectrum of particle predicted by the model.

\subsection{The A factor}

The $A_{\lambda}$, $B_{\mu}$, and $C_{\mu}$ factors can be obtained
by a two step calculation. First one must solve the system
(\ref{E:coef}) by expressing $B$ and $C$ as functions of $A$. Then
one can fix $A$ by requiring $B^{2}+C^{2}=1$, a condition which
comes from the asymptotic behaviour of the Bessel functions.
Following this procedure, and again suppressing factors of $c$ for
notational convenience, we get
\begin{eqnarray}
A_{\nu} 
&=& 
\left.
\frac{W[J_{\nu}(\nl\, \omega_{in} r), N_{\nu}(\nl\, \omega_{in} r)]}
{\sqrt{W[ J_{\nu}(\ng\, \omega_{in} r), N_{\nu}(\nl\, \omega_{in} r)]^2+ 
  W[ J_{\nu}(\ng\, \omega_{in} r), J_{\nu}(\nl\, \omega_{in} r)]^2}}
\right|_{R},
\\ 
B_{\nu} 
&=& 
\left. A_{\nu} 
\frac{W[J_{\nu}(\ng\,\omega_{in}r), N_{\nu}(\nl\, \omega_{in} r)]}
{W[ J_{\nu}(\nl\, \omega_{in} r), N_{\nu}(\nl\, \omega_{in} r) ]}
\right|_{R},
\\
C_{\nu} 
&=& 
\left.
A_{\nu} 
\frac{W[J_{\nu}(\nl\, \omega_{in}r), J_{\nu}(\ng\, \omega_{in} r)]}
{W[J_{\nu}(\nl\, \omega_{in} r), N_{\nu}(\nl\, \omega_{in} r)]}
\right|_{R}.
\label{E:coef2}
\end{eqnarray}
We are mostly interested in the coefficient $A_{\nu}$. This can be
simplified by using a well known formula (Abramowitz-Stegun, page
360 formula 9.1.16) for the (true) Wronskian of Bessel functions of the
first and second kind.
\begin{equation}
W_{true}[J_{\nu}(z), N_{\nu}(z)]=\frac{2}{\pi z}.
\end{equation} 
In our case, taking into account that for our pseudo Wronskian the
derivatives are with respect to $r$ (not with respect to $z$), one
gets for the numerator of $A_{\nu}$:
\begin{equation}
W[J_{\nu}(\nl\, \omega_{in}r), N_{\nu}(\nl\, \omega_{in} r)]_{R}
= \nl\, \omega_{in} 
\frac{2}{\pi (\nl\, \omega_{in} R)}=\frac{2}{\pi R}.
\end{equation}
Hence the $A_{\nu}$ can be written as
\begin{equation}
|A_{\nu}|^{2}= 
\frac{4/(\pi^2 R^2)}{\left. 
W[ J_{\nu}(\ng\,\omega_{in} r), N_{\nu}(\nl\, \omega_{in} r)]^2+
W[ J_{\nu}(\ng\,\omega_{in} r), J_{\nu}(\nl\, \omega_{in} r)]^2 
\right|_{R}}.
\end{equation}
For $\omega\rightarrow \infty$ at $l$ fixed the asymptotic behaviour
is
\begin{equation}
|A_{\nu}|^{2}
\sim
{2\ng\,\nl\over \ng^2+\nl^2 + (\nl^2-\ng^2)\cos(2\ng\,\omega_{in} - 
(\nu+1/2)\pi)}. 
\label{asymp}
\end{equation} 
Numerical plots of  $|A_\nu|^2$  show that it is an oscillating 
function which rapidly reaches this asymptotic form.
%
\begin{figure}[htb]
\vbox{ 
\hfil
\scalebox{\SIZE}{\rotatebox{270}{\includegraphics{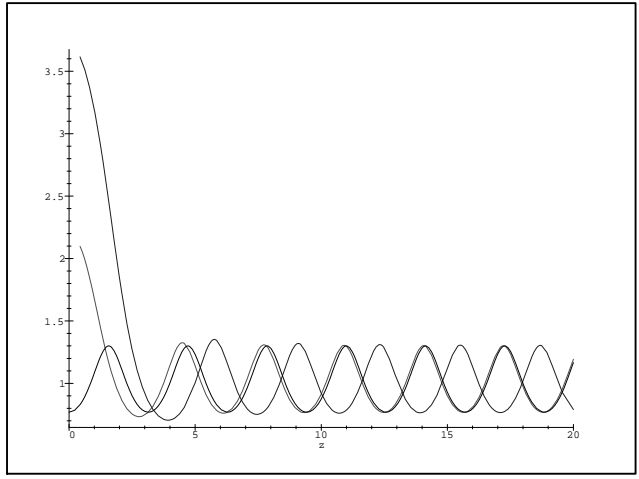}}}
\hfil 
}
\bigskip
\caption{Plot of $|A_{l}|^2$ for $\nu=3/2$ and $\nu=5/2$. (Here we
define $z=\omega R/c$ and put $\ng=1$, $\nl=1.3$.)  The function
rapidly stabilizes to the asymptotic behavior. The dotted curve shows
the behaviour of the asymptotic form for $\nu=3/2$.}
\label{fig:A} 
\end{figure}
%
We shall use this approximation to replace the $A_{\nu}$ factor with its 
mean value for large arguments:
\begin{equation}
|A_{\nu}|^2 \approx {1\over2\pi} 
\int_0^{2\pi} dz 
{2\ng\,\nl\over \ng^2+\nl^2 + (\nl^2-\ng^2)\cos(z)} = 1.
\label{E:app}
\end{equation}
That this approximation is adequate may be checked {\em a posteriori}
by seeing that the Bogolubov coefficients are not noticeably affected.

\subsection{The Pseudo-Wronskian}

Use the simplified notation in which $x = \nl\, \omega_{out} R/c$, $y
= \ng\, \omega_{in} R/c$. In these dimensionless quantities, after
including the approximation equation (\ref{E:app}), and making explicit
the dependence on $R$ and $c$, equation (\ref{E:b2}) takes the form:
\begin{equation}
\label{E:b2-2}
\left|\beta(x,y)\right|^{2}=\frac{R^2}{c^2} 
\frac{(\nl^2-\ng^2)^2}{\nl^2\,\ng^2} 
\left( \frac{y^{2}}{\ng\, x+\nl\, y} 
\right)^2 F(x,y).
\end{equation}
Here $F(x,y)$ is shorthand for the function
\begin{equation}
F(x,y) = \sum_{l=1}^\infty (2l+1)   
{ 
  \left|
  \begin{array}{rr}
  J_{\nu}(x)&J_{\nu}(y)\\
  x\;J^{\prime}_{\nu}(x) & y\;J^{\prime}_{\nu}(y)
  \end{array}
  \right|^{2}
\over
(x^2-y^2)^2
},
\label{E:bint}
\end{equation}
where in this equation the primes now signify derivatives 
with respect to the full arguments ($x$ or $y$).

In order to proceed in our analysis we need now to perform the
summation over angular momentum.  Although the infinite sum is
analytically intractable, there are two reasonable arguments (one
physical and one mathematical) both leading to the conclusion that
suitable truncations of this sum will be enough for our purposes.

The first argument is a physical one and it is based on the maximum 
amount of angular momentum that an outgoing photon may have.
Basically, if one supposes the photons to be produced inside or at most 
on the surface of the bubble, the upper limit for the angular 
momentum  will be the product of the bubble radius
times the maximal ``out" momentum. 
Then one gets: 
\begin{equation}
l_{max}(K) = {R (\hbar K)\over \hbar} = R K.
\label{E:lmax}
\end{equation}
For sonoluminescence $K$ is of order $2\pi/(200 {\rm nm})$. Deciding
the appropriate value of $R$ is more tricky. Since the
sonoluminescence flash occurs at or near the moment of minimum radius
one might wish to use $R_{min}\approx 500 {\rm nm}$ in which case
$l_{max}(K)\approx 15$. Certainly for this choice of $R$ keeping the
first ten or so terms will be sufficient. More conservatively, one
might wish to choose $R$ to be of order $R_{ambient}\approx 4.5 \mu m$
in which case $l_{max}(K) \approx 135$. Keeping this number of terms
in the series is already very unwieldy. Finally, in Schwinger's
original version of the model it is the change in Casimir energy
during the collapse all the way from maximum radius that is relevant,
so perhaps one should use $R_{max} \approx 45 \mu m$. In this case
$l_{max}(K)\approx 1350$, and explicit summation of the series is
prohibitively difficult.  To handle these problems we develop a
semi-analytic approximation to the sum which is sufficient for making
numerical estimates of the spectrum.

This argument can be bolstered by considering the large order
expansion ($\nu \rightarrow \infty$ at fixed $x$) of the Bessel
functions. In this limit one gets \cite{Jeffrey}:
\begin{equation}
\label{E:asymp}
J_{\nu}(x) 
\sim {\frac{1}{\sqrt{2\pi \nu}}}\left(\frac{e x}{2\nu}\right)^{\nu}
\end{equation}
This can be used to obtain the asymptotic form of the 
pseudo-Wronskian appearing in equation (\ref{E:bint}).
\begin{eqnarray}
\tilde W_\nu(x,y)&\equiv&\left|
\begin{array}{rr}
J_{\nu}(x)&J_{\nu}(y)\\
x\;J^{\prime}_{\nu}(x) & y\;J^{\prime}_{\nu}(y)
\end{array}
\right|\\
&=&-\left|
\begin{array}{rr}
J_{\nu}(x)&J_{\nu}(y)\\
x\;J_{\nu+1}(x) & y\;J_{\nu+1}(y)
\end{array}
\right|\\
&\sim& \frac{(x^2-y^2)}{2\pi (\nu)^{1/2} (\nu+1)^{3/2}} 
\left(\frac{xy}{\nu(\nu+1)}\right)^{\nu}  
\left(\frac{e}{2}\right)^{2\nu+1}.
\end{eqnarray}
where we have used the standard recursion relation for the Bessel
functions $J^{\prime}_{\nu}(z)=\nu J_{\nu}(z)-z J_{\nu+1}(z)$.
This indicates that the sum over $\nu$ is convergent: the terms
for which $(xy/\nu^{2})\leq 1$ are suppressed. Since, depending on
one's views as to the appropriate value of $R$,  $x$ and $y$ are
at most of order $15$, $135$, or $1350$ we deduce that the maximal
contribution to the sum comes from a limited number of terms.

Analytically, it is easy to see that the function $F(x,y)$ is finite
along the diagonal and goes smoothly to zero for $x,y \rightarrow
0$.  To proceed to an actual computation of the predicted spectrum
we need to develop an semi-analytic approximate form for this
function by considering separately the behaviour along the diagonal
$x-y=0$ and in the transversal direction $x+y={\rm constant}$.
    
\subsection{Working along the diagonal}

To study in more detail the behaviour of such a function in this zone one 
can perform a Taylor expansion of $F(x,y)$ around $x=y$. 
\begin{eqnarray}
\lim_{x\rightarrow y} \frac{\tilde W_{\nu}(x,y)}{(x-y)} 
&\equiv& \lim_{x\rightarrow y}
\frac{ 
\left|\begin{array}{rr}
J_{\nu}(x)&J_{\nu}(y)\\
x\;J^{\prime}_{\nu}(x) & y\;J^{\prime}_{\nu}(y)
\end{array}
\right|}
{(x-y)}\\
&=& \lim_{x\rightarrow y}
\frac{
\left|\begin{array}{ll}
J_{\nu}(x)& \hphantom{x\;} J_{\nu}(x)+(x-y)J^{\prime}_{\nu}(x)\\
x\;J^{\prime}_{\nu}(x) & 
x\;J^{\prime}_{\nu}(x)+(x-y)
[J^{\prime}_{\nu}(x)+x\;J^{\prime\prime}_{\nu}(x)] 
\end{array} 
\right|}   
{(x-y)}\\
&=& 
\left|\begin{array}{rl}
J_{\nu}(x)&J^{\prime}_{\nu}(x)\\
x\;J^{\prime}_{\nu}(x) & 
J^{\prime}_{\nu}(x)+x\;J^{\prime\prime}_{\nu}(x)
\end{array} 
\right| \\
&=&
J_{\nu}(x)[J^{\prime}_{\nu}(x)+x\;J^{\prime\prime}_{\nu}(x)]-
x\;{J^{\prime}_{\nu}(x)}^{2}.
\end{eqnarray}
The derivatives can be eliminated by using the well known recursion 
relations.
\begin{eqnarray} 
\lim_{x\rightarrow y} \frac{\tilde W_{\nu}(x,y)}{(x-y)}
&=&
J_{\nu}(x)
\left[\frac{(\nu^{2}-x^{2})}{x}\right]
-x\;\left[\frac{\nu}{x} J_{\nu}(x)-J_{\nu+1}(x)\right]^{2}\\
&=& 
2 \nu 
J_{\nu}(x)J_{\nu+1}(x)-x\;\left[J_{\nu}^{2}(x)+J_{\nu+1}^{2}(x)\right].
\label{E:diatrm} 
\end{eqnarray} 
For sake of simplicity we shall use an equivalent form of equation
(\ref{E:diatrm}) where lower order Bessel function appear
\begin{equation}
\lim_{x\rightarrow y} \frac{\tilde W_{\nu}(x,y)}{(x-y)}=2 \nu 
J_{\nu}(x)J_{\nu-1}(x)-x\;\left[J_{\nu}^{2}(x)+J_{\nu-1}^{2}(x)\right].
\label{E:dia}
\end{equation} 
This result shows that, as expected, each term of $F(x,x)$ is 
finite along the diagonal and equal to zero at $x=y=0$.
Moreover 
\begin{equation}
D(x)\equiv F(x,x)=\sum_{l=1}^{\infty}(2l+1) 
{\left\{
(2l+1) J_{l+1/2}(x) J_{l-1/2}(x) 
-x \; \left[J_{l+1/2}^{2}(x)+J_{l-1/2}^{2}(x)\right]
\right\}^2
\over
4x^{2}}. 
\end{equation} 
This sum can easily be checked to be convergent for fixed $x$. [Use
equation (\ref{E:asymp}).] With a little more work it can be shown that
\[
\lim_{x\to\infty} D(x) = {1\over2\pi^2}.
\]
The truncated function obtained after summation over the first few
terms (say the first ten or so terms) is a long and messy combination
of trigonometric functions that can however be easily plotted and
approximated in the range of interest. Due to numerical artifacts,
the function is not controllable near the origin, fortunately we
have analytic information in that region --- the function is very near
to zero in the range $(0,1)$ for both ``out'' and ``in'' frequencies,
and can be approximated by zero without any undue influence on the
numerical results.  A semi-analytical study led us to the approximate
form of $D(x)$
\begin{equation}
D(x) \approx \Theta(x-1)\;\frac{1}{2\pi^{2}}
\frac{2 (x-1)^{2}}{3+2(x-1)^{2}}.
\end{equation}
A confrontation between the two curves in the range of interest is given 
in the figure below.
%

\begin{figure}[htb]
\vbox{ 
\hfil
\scalebox{\SIZE}{\rotatebox{270}{\includegraphics{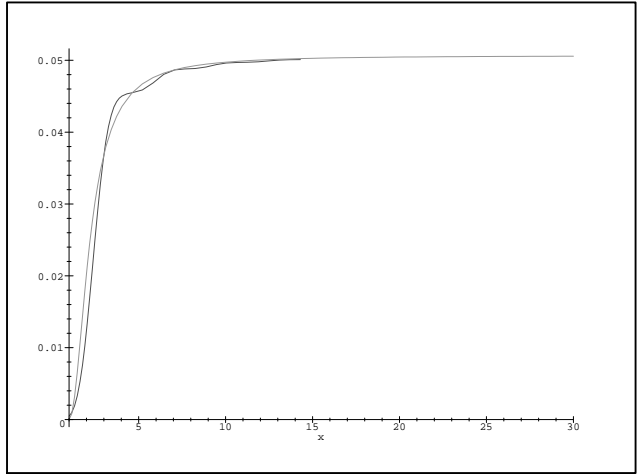}}}
\hfil 
}
\bigskip
\caption{Plot of the exact $D(x)$ against its approximated form in the 
range $1<x<30$}  
\label{fig:diagapp}
\end{figure}

\subsection{The factorization approximation}

To numerically perform the integrals needed to do obtain the spectrum
it is useful to note the approximate factorization property
\begin{equation}
F(x,y) \approx F\left({x+y\over2},{x+y\over2}\right) \; 
G\left(\frac{x-y}{2}\right). 
\end{equation}
That is: to a good approximation $F(x,y)$ is given by its value
along the nearest part of the diagonal, multiplied by a universal
function of the distance away from the diagonal. A little experimental
curve fitting is actually enough to show that to a good approximation
\begin{equation}
F(x,y) \approx  D\left({x+y\over2}\right) \; 
{\sin^2(\pi[x-y]/4)\over (\pi[x-y]/4)^2} .
\end{equation}
{From} the plot we show below it is easy to check that the function
$F(x,y)$ is quite well fitted by our approximation. We feel important
to stress that this is approximation is based on numerical
experimentation, and is not an analytically-driven approximation.
(In the infinite volume case we know that $F(x,y) \to (constant)
\times \delta(x-y)$, {\em cf.} equation (\ref{E:infinite-volume}).  The
effect of finite volume is effectively to ``smear out'' the delta
function. The combination $\sin^2(x)/(\pi x^2)$ is one of the
standard approximations to the delta function.) Our approximation
is quite good everywhere except for values of $x$ and $y$ near the
origin (less than 1) where the contribution of the function to the
integral is very small.
%

\begin{figure}[htb]
\vbox{ 
\hfil
\scalebox{\SIZE}{\rotatebox{270}{\includegraphics{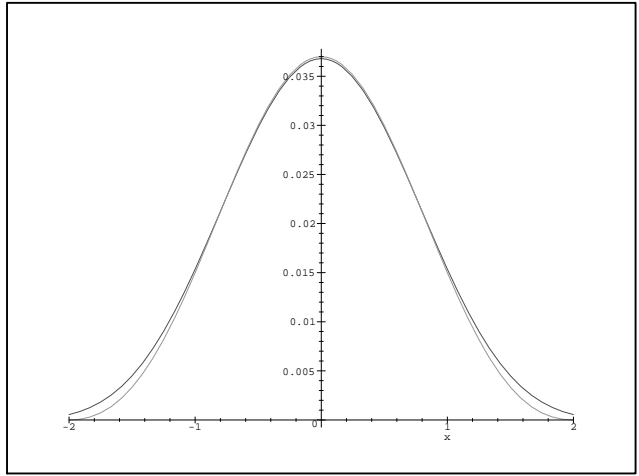}}}
\hfil 
}
\bigskip
\caption{Transverse fit: An orthogonal slice of $F(x,y)$ intersecting
the diagonal at $(x,y)=(3,3)$. Here $F(3+z,3-z)$ is plotted  in
comparison with $[\sin^2(\pi z/2)]/(\pi z/2)^2 $. }
\label{fig:trsvapp1}
\end{figure}
%
\begin{figure}[htb]
\vbox{ 
\hfil
\scalebox{\SIZE}{\rotatebox{270}{\includegraphics{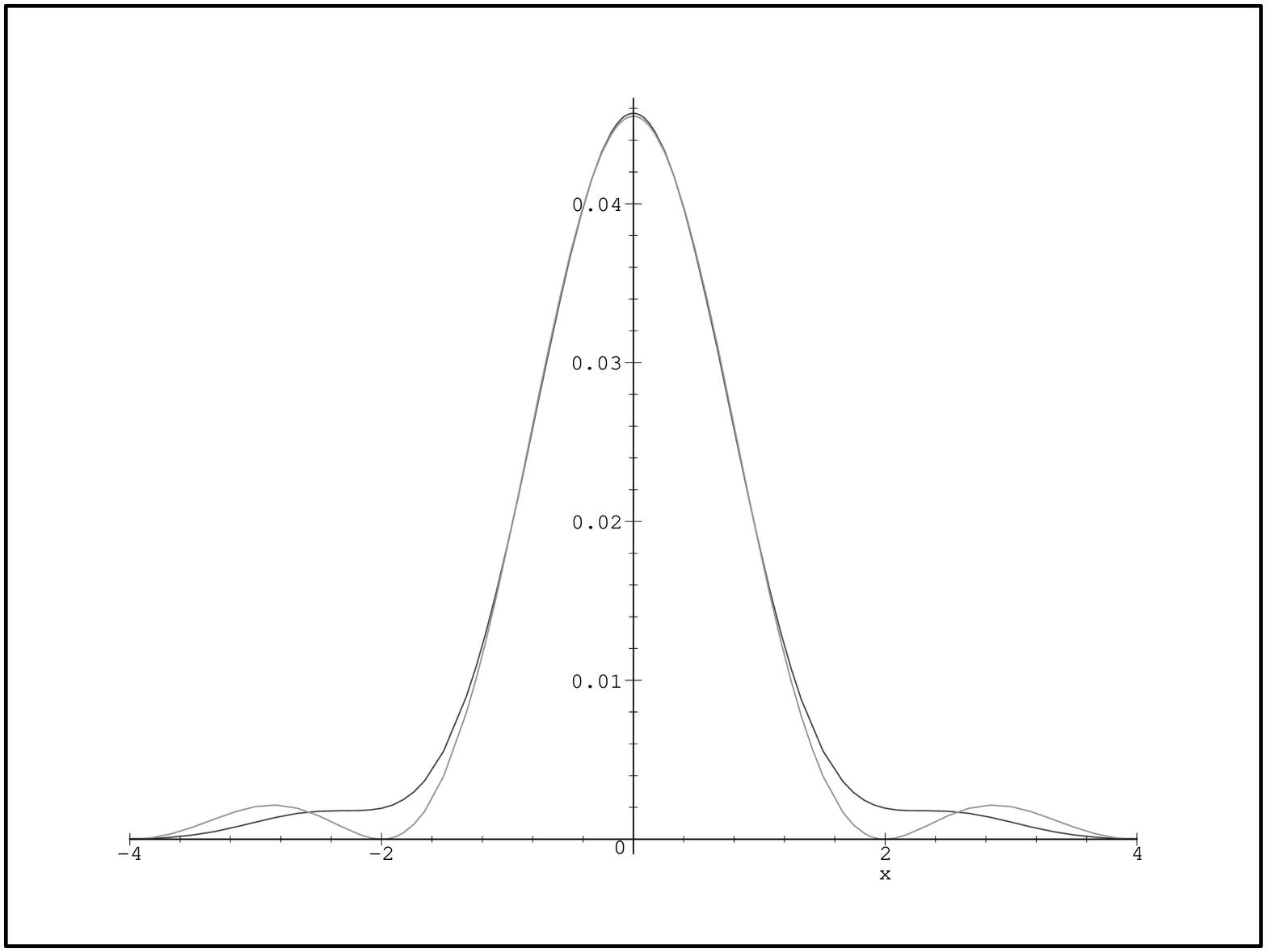}}}
\hfil 
}
\bigskip
\caption{Transverse fit: An orthogonal slice of $F(x,y)$ intersecting
the diagonal at $(x,y)=(5,5)$. Here $F(5+z,5-z)$ is plotted in
comparison with $[\sin^2(\pi z/2)]/(\pi z/2)^2 $. }
\label{fig:trsvapp2}
\end{figure}

\begin{figure}[htb]
\vbox{ 
\hfil
\scalebox{\SIZE}{\rotatebox{270}{\includegraphics{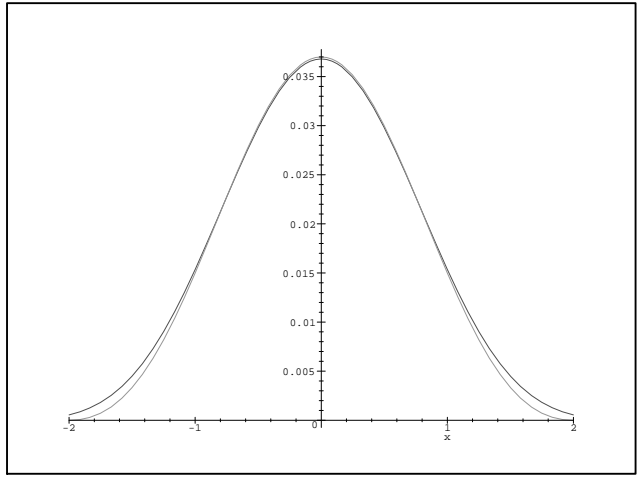}}}
\hfil 
}
\bigskip
\caption{Transverse fit: An orthogonal slice of $F(x,y)$ intersecting
the diagonal at $(x,y)=(10,10)$. Here $F(10+z,10-z)$ is plotted in
comparison with $[\sin^2(\pi z/2)]/(\pi z/2)^2 $. }
\label{fig:trsvapp3}
\end{figure}
%

\subsection{The spectrum: numerical evaluation}

We have now transformed the function $F(x,y)$ into an easy to handle
product of two functions
\begin{equation}
F(x,y) \approx  \Theta(x+y-2) \;
\frac{1}{2\pi^{2}} \;
\frac{(x+y-2)^{2}}{6+(x+y-2)^{2}} \;
{\sin^2(\pi[x-y]/4)\over(\pi[x-y]/4)^2}. 
\label{fapp}
\end{equation}
We exhibit tridimensional graphs for both the exact (apart from the
approximation of truncating the sum at a finite $l$) and approximate
forms of the function $F(x,y)$.  We have chosen the case of $R=0.5\mu
{\rm m}$ (corresponding to $y_{max}=2.5$ as previously explained).
%

\begin{figure}[htb]
\vbox{ 
\hfil
\scalebox{\SIZE}{\rotatebox{270}{\includegraphics{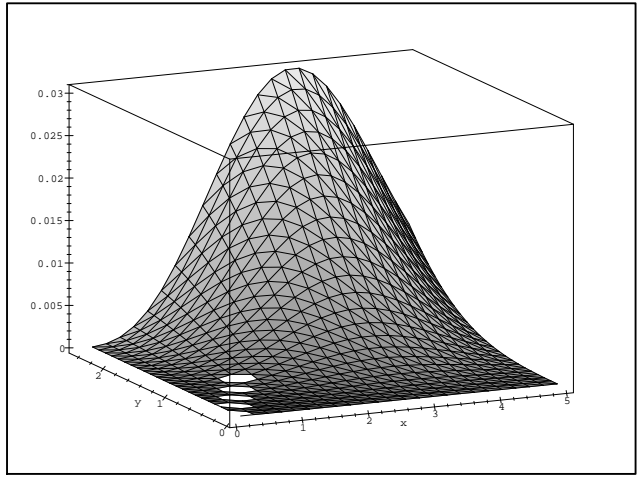}}}
\hfil 
}
\bigskip
\caption{Plot of the exact $F(x,y)$ in the range $0<x<5$, $0<y<2.5$}
\label{fig:th}
\end{figure}

\begin{figure}[htb]
\vbox{ 
\hfil
\scalebox{\SIZE}{\rotatebox{270}{\includegraphics{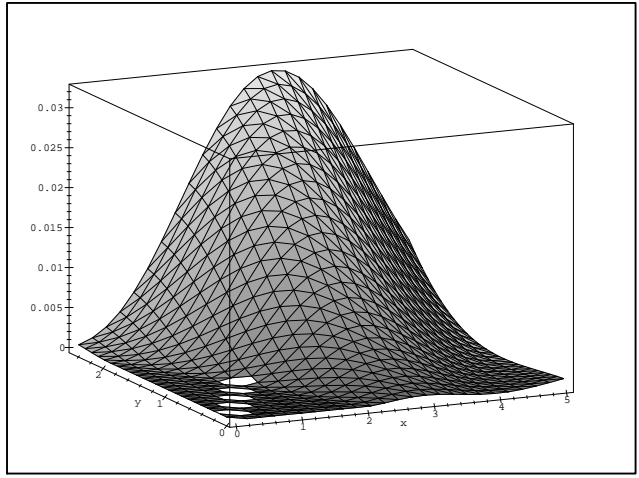}}}
\hfil 
}
\bigskip
\caption{Plot of the approximated $F(x,y)$ in the range $0<x<5$,
$0<y<2.5$}
\label{fig:thexp}
\end{figure}

A numerical study of the error due to the replacement of $F(x,y)$ with
its approximated form equation (\ref{fapp}), leads to an upper limit
of $20\%$ error in the total energy emitted.

The dimensionless spectrum, based on equations (\ref{E:b2}) and
(\ref{E:b2-2}), is
\begin{equation}
\label{E:dimensionless-spectrum-2}
{dN\over dx} 
=\frac{(\nl^2-\ng^2)^2}{\nl^3\,\ng^3} 
\int_0^{RK} 
\left( \frac{y^{2}}{\ng\, x+\nl\, y} \right)^2 
D\left({x+y\over2}\right) 
{\sin^2(\pi[x-y]/4)\over(\pi[x-y]/4)^2} dy.
\end{equation}
As a consistency check, the infinite volume limit is equivalent to
making the formal replacements
\begin{equation}
{\sin^2(\pi[x-y]/4)\over(\pi[x-y]/4)^2} \to 4 \delta(x-y),
\end{equation}
and
\begin{equation}
D\left({x+y\over2}\right) \to {1\over2\pi^2}.
\end{equation}
Doing so, equation (\ref{E:dimensionless-spectrum-2}) reduces to
equation (\ref{E:dimensionless-spectrum-1}) up to an overall factor
[$4/\pi$] of order one. The correct dependence on refractive index and
correct power-law behaviour for the spectrum are recovered, and the
overall order one factor is merely a reflection of the crudity of the
cutoff in angular momentum used in deriving
(\ref{E:dimensionless-spectrum-1}).

With this consistency check out of the way, it is now possible to
perform the integral with respect to $y$ to estimate the spectrum
for finite volume.  For definiteness we set $\nl = 1.3$ and $\ng=1.0$,
put $K=2\pi/(200 nm)$, and pick $R=0.5\mu {\rm m}$ (corresponding to
$y_{max}=15$).  We integrate from $y=0$ to $y=15$ and plot the
resulting spectrum from $x=0$ to $x=18$.

\begin{figure}[htb]
\vbox{ 
\hfil
\scalebox{\SIZE}{\rotatebox{270}{\includegraphics{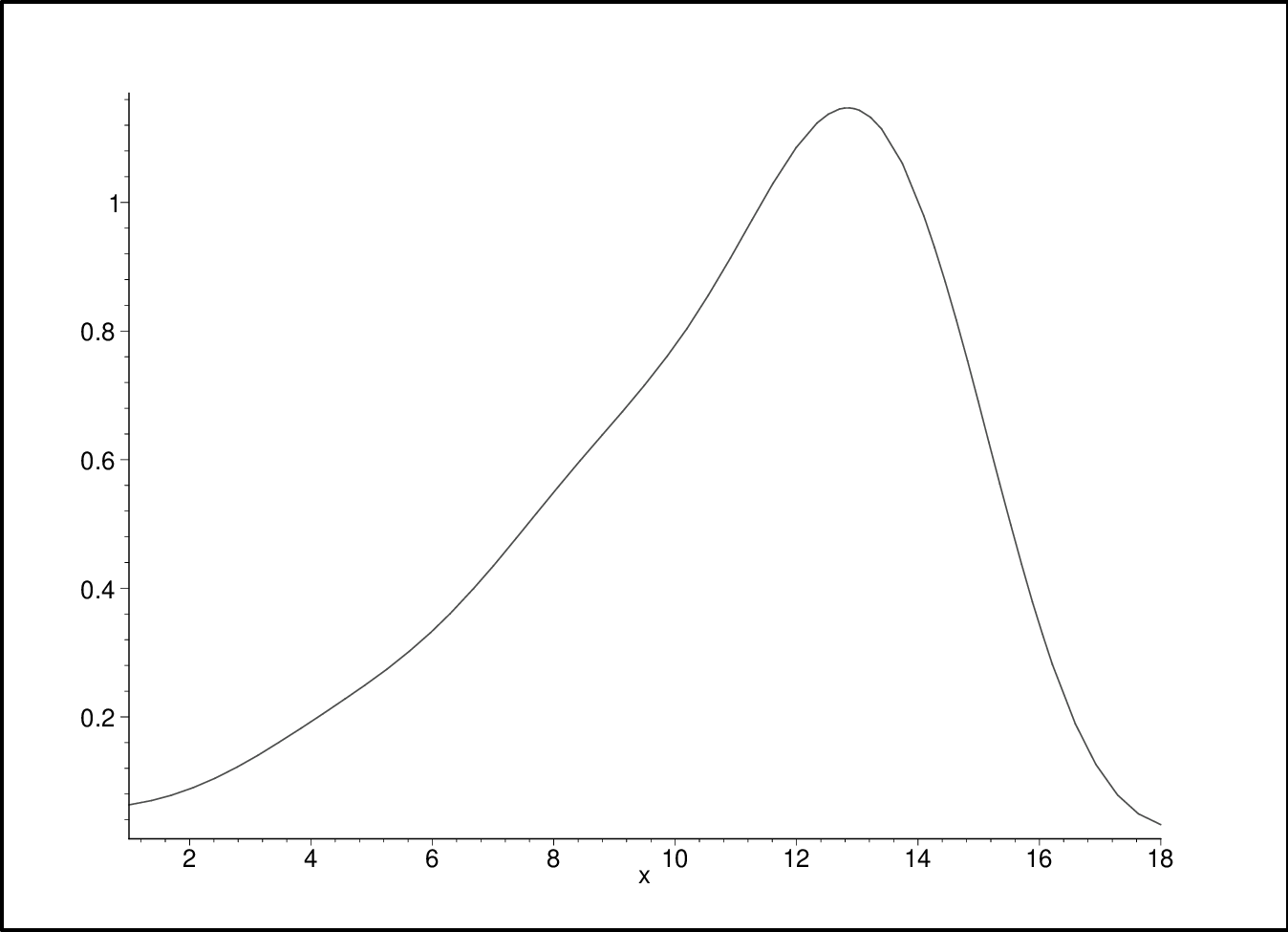}}}
\hfil 
}
\bigskip
\caption{Spectrum obtained by the approximated Bogolubov coefficient
for $R=0.5\mu {\rm m}$ corresponding to $y_{max}=15$.  
We integrate from $y=0$ to $y=15$ and plot the resulting spectrum from 
$x=0$ to $x=18$.} 
\label{fig:sp1} 
\end{figure}

One can also ask what sort of result one would get if instead we
pick a much larger value of $R$, say $R=5 \mu m$, corresponding to
the bubble at equilibrium radius. In this case the approach towards
the Schwinger (infinite volume result) result is much closer. We
now have $y_{max}=135$.  We integrate from $y=0$ to $y=135$ and plot
the resulting spectrum from $x=0$ to $x=140$. For comparison we plot it
together with equation (\ref{E:dimensionless-spectrum-1}) which is
Schwinger's naive model (the re-scaled infinite volume limit).

\begin{figure}[htb]
\vbox{ 
\hfil
\scalebox{\SIZE}{\rotatebox{270}{\includegraphics{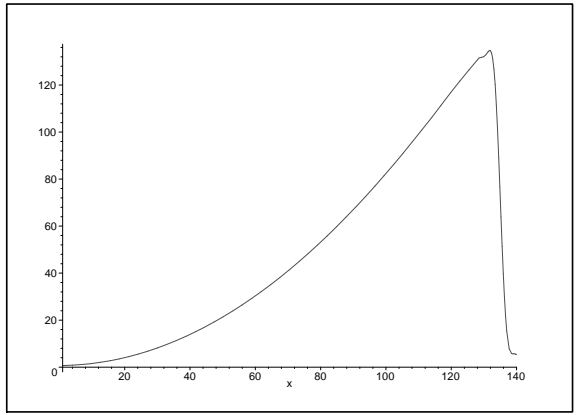}}}
\hfil 
}
\bigskip
\caption{Spectrum obtained by the approximated Bogolubov coefficient
for $R=5\mu {\rm m}$ corresponding to $y_{max}=135$.
We integrate from $y=0$ to $y=135$ and plot the resulting spectrum from
$x=0$ to $x=140$.}
\label{fig:sp2}
\end{figure}

The case corresponding to the bubble at maximum radius, $R=50 \mu m$,
requires a range of integration too large for standard numerical
plotting. In any case the graph will only be a replica of the previous
one on a larger scale.

\section{Discussion}

The lessons we have learned from this test calculation are:

---(1) The model proves (in an indirect way) that the Casimir energy
produced via the bubble collapse includes (in the large $R$ limit) a
term proportional to the volume (actually to the volume over which the
refractive index changes). In the case of a truly dynamical model one
expects that the energy of the photons so created will be provided by
other sources of energy ({\em e.g.}, the sound wave), nevertheless the
presence of a volume contribution appears unavoidable.

---(2) The present model is still unable to fully fit other
experimental features of sonoluminescence. For example it provides
maximal photon release at maximum expansion. Barber {\em et
al.}~\cite{Physics-Reports} point out that in Schwinger's original
model the main production of photons may be expected when the the rate
of change of the volume is maximum, which is experimentally found to
occur near the maximum radius. In contrast the emission of light in
sonoluminescence is experimentally found to occur near the point of
minimum radius, where the rate of change of area is maximum. {\em All
else being equal}, this would seem to indicate a surface dependence
and might be interpreted as a true weakness of the dynamical Casimir
explanation of sonoluminescence.  

In fact we have shown elsewhere~\cite{letter,qed1,qed2,2gamma} that
the situation is considerably more complex than might naively be
thought. It is important to stress that what Schwinger proposed was
clearly {\em only} a first estimate of the vacuum energy, which was in
principle viable as the basis for a model, and {\em not} a fully
dynamical model.  Schwinger was fully aware of this in his papers.

A fully dynamical calculation is required in order to deal with these
issues, and the experimental data give remarkable suggestions about
the plausible directions for theoretical developments within the
framework of the dynamic Casimir effect.  In particular, one of the
key features of photon production by a space-dependent and
time-dependent refractive index is that for a change occurring on a
timescale $\tau$, the amount of photon production is exponentially
suppressed by an amount $\exp(-\omega\tau)$. In~\cite{qed1} we have
provided a specific model that exhibits this behaviour, and argued
that the result is in fact generic.  The importance for
sonoluminescence is that the experimental spectrum is {\em not\,}
exponentially suppressed at least out to the far ultraviolet.
Therefore any mechanism of Casimir-induced photon production based on
an adiabatic approximation is destined to failure: Since the
exponential suppression is not visible out to $\omega\approx10^{15} \;
\hbox{Hz}$, it follows that {\em if\,} sonoluminescence is to be
attributed to photon production from a time-dependent dielectric
bubble ({\em i.e.}, the dynamical Casimir effect), {\em then} the
timescale for change in the refractive index must be of order of a
{\em femtosecond}.  Thus any Casimir--based model has to take into
account that {\em it is no longer the collapse from $R_{\mathrm max}$
to $R_{\mathrm min}$ that is important}. One has to divorce the change
in refractive index from direct coupling to the bubble wall motion,
and instead ask for a rapid change in the refractive index of the
entrained gases as they are compressed down to their van der Waals
hard core~\cite{qed1,qed2}.  We stress that this conclusion, though it
moves away from the original Schwinger proposal, is still firmly
within the realm of the dynamic Casimir effect approach to
sonoluminescence. The fact is that the present work shows clearly that
a viable Casimir ``route'' to sonoluminescence cannot avoid a ``fierce
marriage'' between QFT and features related to condensed matter
physics.

\section{Conclusions}

The present calculation unambiguously verifies that a sudden change in
radius of a dielectric bubble causes a change in the Casimir energy
that is, as predicted by Schwinger~\cite{Sc4,CMMV1,CMMV2,MV},
converted into real photons with a phase space spectrum.  As far as
sonoluminescence is concerned, we have also explained why such a
change {\em must} be sudden in order to fit the experimental data.
This leads us to propose a somewhat different model of
sonoluminescence based on the dynamical Casimir effect, a model
focussed on the actual dynamics of the refractive index (as a function
of space and time), and not just of the bubble boundary. (In
Schwinger's original approach the refractive index changes only due to
motion of the bubble wall.)  In summary, provided the sudden
approximation is valid, changes in the refractive index will lead to
efficient conversion of zero point fluctuations into real photons.
Trying to fit the details of the observed spectrum in sonoluminescence
then becomes an issue of building a robust model of the refractive
index of both the ambient water and the entrained gases as functions
of frequency, density, and composition. Only after this prerequisite
is satisfied will we be in a position to develop a more complex
dynamical model endowed with adequate predictive power.
                                                            
In light of these observations we think that one can also derive a
general conclusion about the long standing debate on the actual value
of the static Casimir energy and its relevance to sonoluminescence:
Sonoluminescence is not directly related to the {\em static} Casimir
effect.  The static Casimir energy is at best capable of giving a
crude estimate for the energy budget in sonoluminescence.

\section*{Acknowledgements}

This research was supported by the Italian Ministry of Science (DWS,
SL, and FB), and by the US Department of Energy (MV). MV particularly
wishes to thank SISSA (Trieste, Italy) for hospitality during closing
phases of this research. DWS and SL wish to thank E.~Tosatti for
useful discussion. SL wishes to thank M.~Bertola and B.~Bassett for
comments and suggestions.



\begin{thebibliography}{99}
\bibitem[\dagger]{} E-mail: liberati@sissa.it
\bibitem[\P]{} E-mail: visser@kiwi.wustl.edu
\bibitem[\ddag]{} E-mail: belgiorno@mi.infn.it
\bibitem[\S]{} E-mail: sciama@sissa.it
\bibitem{Sc1} 
J. Schwinger,
Proc. Nat. Acad. Sci. {\bf 89}, 4091--4093 (1992).
\bibitem{Sc2} 
J. Schwinger,
Proc. Nat. Acad. Sci. {\bf 89}, 11118--11120 (1992).
\bibitem{Sc3} 
J. Schwinger,
Proc. Nat. Acad. Sci. {\bf 90}, 958--959 (1993).
\bibitem{Sc4} 
J. Schwinger,
Proc. Nat. Acad. Sci. {\bf 90}, 2105--2106 (1993).
\bibitem{Sc5} 
J. Schwinger,
Proc. Nat. Acad. Sci. {\bf 90}, 4505--4507 (1993).
\bibitem{Sc6} 
J. Schwinger,
Proc. Nat. Acad. Sci. {\bf 90}, 7285--7287 (1993).
\bibitem{Sc7} 
J. Schwinger,
Proc. Nat. Acad. Sci. {\bf 91}, 6473--6475 (1994).
\bibitem{Physics-Reports} 
B.P. Barber, R.A. Hiller, R. L\"{o}fstedt, S.J. Putterman
Phys. Rep. {\bf 281}, 65-143 (1997).
\bibitem{letter}
S. Liberati, M. Visser, F. Belgiorno, and D.W. Sciama,
{\em Sonoluminescence: Bogolubov coefficients for the QED vacuum
of a collapsing bubble}, quant-ph/9805023, 
to appear in Physical Review Letters.
\bibitem{qed1}
S. Liberati, M. Visser, F. Belgiorno, and D.W. Sciama,
{\em Sonoluminescence as a QED vacuum effect. I: Physical Scenario},
quant-ph/9904013.    
\bibitem{qed2}
S. Liberati, M. Visser, F. Belgiorno, and D.W. Sciama,
{\em Sonoluminescence as a QED vacuum effect. II: Finite Volume Effects},
quant-ph/9905034.         
\bibitem{2gamma}
F. Belgiorno, S. Liberati, M. Visser, and D.W. Sciama,
{\em Sonoluminescence: Two-photon correlations as a test for thermality},
quant-ph/9904018.
\bibitem{Eberlein1}
C. Eberlein,
{\em Sonoluminescence as quantum vacuum radiation},
Phys. Rev. Lett. {\bf 76}, 3842 (1996).
quant-ph 9506023
\bibitem{Eberlein2} 
C. Eberlein, 
{\em Theory of quantum radiation observed as sonoluminescence},
Phys. Rev. {\bf A 53}, 2772 (1996). 
quant-ph 9506024
\bibitem{Eberlein3}
C. Eberlein,
{\em Sonoluminescence as quantum vacuum radiation (reply to comment)},
Phys. Rev. Lett. {\bf 78}, 2269 (1997).
quant-ph/9610034 
\bibitem{CMMV1} 
C. E. Carlson, C. Molina--\Paris,
J. \Perez--Mercader, and M. Visser,
Phys. Lett. {\bf B 395}, 76-82 (1997). hep-th/9609195 
\bibitem{CMMV2} 
C. E. Carlson, C. Molina--\Paris,
J. \Perez--Mercader, and M. Visser,
Phys. Rev. {\bf D56}, 1262 (1997). hep-th/9702007. 
\bibitem{MV} C. Molina--\Paris\ and M. Visser,
Phys. Rev. {\bf D56}, 6629 (1997). hep-th/9707073.
\bibitem{M95} K. Milton, 
{\em Casimir energy for a spherical cavity in a dielectric: toward a 
model for Sonoluminescence?}, in
{\em Quantum field theory under the influence of external conditions},
edited by M. Bordag, (Tuebner Verlagsgesellschaft, Stuttgart, 1996),
pages 13--23.  See also hep-th/9510091. 
\bibitem{M96} K. Milton and J. Ng, 
{\em Casimir energy for a
spherical cavity in a dielectric: Applications to Sonoluminescence},
hep-th/9607186.
\bibitem{M97} K. Milton and J. Ng,
{\em Observability of the bulk Casimir effect: Can the dynamical Casimir 
effect be relevant to Sonoluminescence ?},
hep-th/9707122.
\bibitem{Bateman} 
H. Bateman,
{\em Higher Transcendental Functions},
Vol II, (McGraw-Hill, New York, 1953).
\bibitem{Jeffrey} 
A. Jeffrey,
{\em Handbook of Mathematical Formulas and Integrals}, 
page 219, (Academic Press, San Diego, 1995).
\end{thebibliography}
\end{document}